\documentclass[a4paper,11pt]{article}
\pdfoutput=1 % if your are submitting a pdflatex (i.e. if you have
\usepackage{jinstpub} % for details on the use of the package, please
\title{\boldmath Measurements with silicon detectors at extreme neutron fluences$^{*}$ \note[*]{Work performed in the framework of the CERN-RD50 collaboration.} }

\author[a,1]{I. Mandi\' c,\note{Corresponding author.}}
\author[a]{V. Cindro,}
\author[a]{A. Gori\v sek,}
\author[a]{B. Hiti,}
\author[a]{G. Kramberger,}
\author[a,b]{M. Miku\v z,}
\author[a]{M. Zavrtanik,}
\author[a]{P. Skomina,}
\author[c]{S. Hidalgo,}
\author[c]{G. Pellegrini}
\affiliation[a]{Jo\v zef Stefan Institute, Ljubljana, Slovenia}
\affiliation[b]{Department of Physics, University of Ljubljana, Slovenia}
\affiliation[b]{CNM-IMB-CSIC, Barcelona, Spain}

\emailAdd{igor.mandic@ijs.si}

\abstract{Thin pad detectors made from 75 $\mu$m thick epitaxial silicon on low resistivity substrate were irradiated with reactor neutrons to fluences from 2.5$\times 10^{16}$ n/cm$^2$ to 1$\times 10^{17}$ n/cm$^2$. Edge-TCT measurements showed that the active detector thickness is limited to the epitaxial layer and does not extend into the low resistivity substrate even after the highest fluence.
Detector current was measured under reverse and forward bias. The forward current was higher than the reverse at the same voltage but the difference gets smaller with increasing fluence. Rapid increase of current (breakdown) above ~ 700 V under reverse bias was observed. An annealing study at 60$^\circ$C was made to 1200 minutes of accumulated annealing time.
It showed that the reverse current anneals with similar time constants as measured at lower fluences. A small increase of forward current due to annealing was seen.
Collected charge was measured with electrons from $^{90}$Sr source in forward and reverse bias configurations.
Under reverse bias the collected charge increased linearly with bias voltage up to 6000 electrons at 2.5$\times 10^{16}$ n/cm$^2$ and 3000 electrons at 1$\times 10^{17}$ n/cm$^2$. Rapid increase of noise was measured
above $\sim$ 700 V reverse bias due to breakdown resulting in worse S/N ratio. At low bias voltages slightly more charge is measured under forward bias compared to reverse. However better S/N is achieved under reverse bias.  
Effective trapping times were estimated from charge collection measurements
under forward bias showing that at high fluences they are much longer than values extrapolated from low fluence measurements - at 1$\times 10^{17}$ n/cm$^2$ a factor of 6 larger value was measured.
}


\begin{document}
\maketitle
\flushbottom

\section{Introduction}
\label{sec:intro}

In this paper we report about measurements with silicon detectors irradiated up to 10$^{17}$ n/cm$^2$ 1 MeV neutron equivalent fluence in silicon. Such radiation levels are expected in hadron colliders planned for the post LHC era. One of the proposals is FCC-hh - the Future Circular Collider for hadrons, a 100 km long
synchrotron at CERN. FCC will accelerate protons up to energies of 50 TeV per beam allowing 100 TeV collisions at very high rate
\cite{FCC_cdr}, generating an extreme radiation environment for the tracking detectors near the interaction point \cite{riegler}.
Estimates of FCC hadron fluences in the innermost tracking layers after integrated luminosity of
30 ab$^{-1}$ approach 10$^{18}$ n$_{eq}$/cm$^2$. This is several times harsher than, for example, in the upgraded ATLAS
experiment at HL-LHL where the innermost (exchangeable) pixel layer in ATLAS must withstand a fluence of 2$\cdot$10$^{16}$ n/cm$^2$ \cite{radlevel, pixelTDR}, which is the highest radiation level where detectors will operate in near future particle physics experiments.

Not many measurements at fluences in the range of 10$^{17}$ n/cm$^2$ have been published so far.  Measurements of charge collection with $^{90}$Sr source up to 1.6$\cdot$10$^{17}$ n/cm$^2$ 
are reported in \cite{spagheti}. Preliminary Edge-TCT studies with strip detectors irradiated up to 10$^{17}$ n/cm$^2$ presented at workshops \cite{mikuz-trento, mikuzFCC} showed that information about carrier mobility can be extracted from these measurements. But obviously many more studies are necessary to understand performance of silicon detectors in such high radiation environment and to extract parameters needed for reliable modeling of detector performance at future colliders.
In this work we write about a systematic set of measurements with thin silicon pad detectors irradiated up to 10$^{17}$ n/cm$^2$.

\section{Samples}

It is clear that at high fluences detectors should be thin and operated at high bias voltages to increase charge collection.
At high fluences thinner detectors outperform thicker ones at same bias voltage \cite{gianluigi_thin}. The reason is the larger
electric field and consequently longer drift distance, while the effect of more charge generated in a thicker detector by the charged particle
crossing is offset by drift in the lower weighting field.

\begin{figure}[htbp]
  \centering % \begin{center}/\end{center} takes some additional vertical space

\includegraphics[width=.4\textwidth]{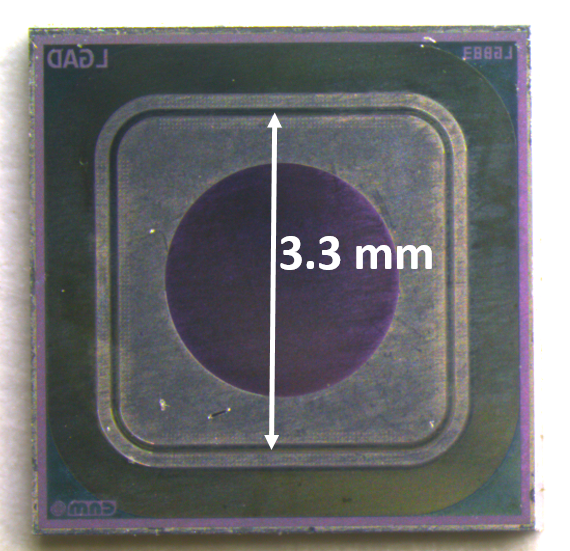}

\caption{\label{detphoto} Photo of the detector used in this study. The dimension of the collection electrode with a hole in the metallization is indicated. The electrode is surrounded by a guard ring and by a p-stop implant. } 
\end{figure}

For this study a set of thin 3 mm x 3 mm pad detectors with one guard ring were used (Fig. \ref{detphoto}). Detectors were produced by CNM, Spain \cite{cnm} on a 75 $\mu$m thick p-type epitaxial layer with resistivity of $\sim$ 100 $\Omega$cm grown on 500 $\mu$m thick low resistivity (6$\cdot 10^{-3} \Omega$cm) Cz substrate. The detectors are actually Low Gain Avalanche Detectors (LGAD) \cite{lgad}
so they have a thin (few  $\mu$m) p-type multiplication layer implanted below the n$^+$ layer of the collection electrode surrounded by junction termination extension (JTE) structure allowing operation up to high bias voltages. The gain of these devices before irradiation was low, slightly above 1 \cite{carula}, and this gain was lost by heavy irradiation because of the initial acceptor removal \cite{removal} so they behaved like standard thin p-type pad detectors. They were used because 75 $\mu m$ is the right thickness to ensure high electric field at reachable bias voltages and at this thickness E-TCT measurements are still feasible.

\section{Irradiation and measurements}

Detectors were irradiated with neutrons in the TRIGA reactor in Ljubljana \cite{Reactor1,Reactor2} to three fluences: 2.5$\cdot$10$^{16}$, 5.7$\cdot$10$^{16}$ and 1$\cdot$10$^{17}$ n$_{eq}$/cm$^2$ (1 MeV neutron equivalent),two devices per fluence. Three
irradiation channels were used for this irradiation, each with different neutron flux (see \cite{Reactor1}) so that irradiation
time for the three fluences was about 4 hours. After irradiation detectors were kept at temperature of -20$^\circ$C. The uncertainty on the fluences is about 10\%. 

\subsection{Edge-TCT}

After irradiation, Edge-TCT \cite{Edge-TCT} measurements were performed mainly to cross check the thickness of the active region. Although the devices were fabricated on a substrate with very low initial resistivity, this check was made to exclude the possibility that the electric field extends into the substrate after heavy
irradiation.  As explained in \cite{removal}, irradiation of p-type silicon with hadrons may lead to lowering the effective space charge concentration in certain range of fluences. The reason is electrical neutralization of initial acceptors - boron in this case - through the interaction with radiation induced defects in silicon. The process is called initial acceptor removal. As a consequence of this process there is a possibility that significant part of initially very low resistivity Cz substrate would be depleted after heavy neutron irradiation.

Measurements were made on the Edge-TCT setup
produced by Particulars \cite{Particulars}. Pad detectors are not ideal for Edge-TCT measurements because laser beam is running under 3 mm of 
readout electrode in the direction of the laser beam. As the laser light propagates through the silicon the beam spot diameter increases beyond the beam waist due to diffraction \cite{paralel} and the depth of charge creation is not defined as well as in strip detectors. This effect worsens the resolution of measurement although it is expected to be somewhat less pronounced
after high irradiation because of the increased laser light absorption \cite{scharf}. In addition, because of conical shape of laser beam, illumination of the top side of the detector through the opening in metallization and reflections of laser light from the top electrode metal may obscure the measurement. To avoid these uncertainties, Edge-TCT measurements were made by observing the induced current from the guard ring - the n-doped ring
surrounding the main electrode without the multiplication layer (see Fig. \ref{detphoto}) instead of the main electrode.

\begin{figure}[htbp]
  \centering % \begin{center}/\end{center} takes some additional vertical space

 \begin{tabular}{ c c c }
\includegraphics[width=.3\textwidth]{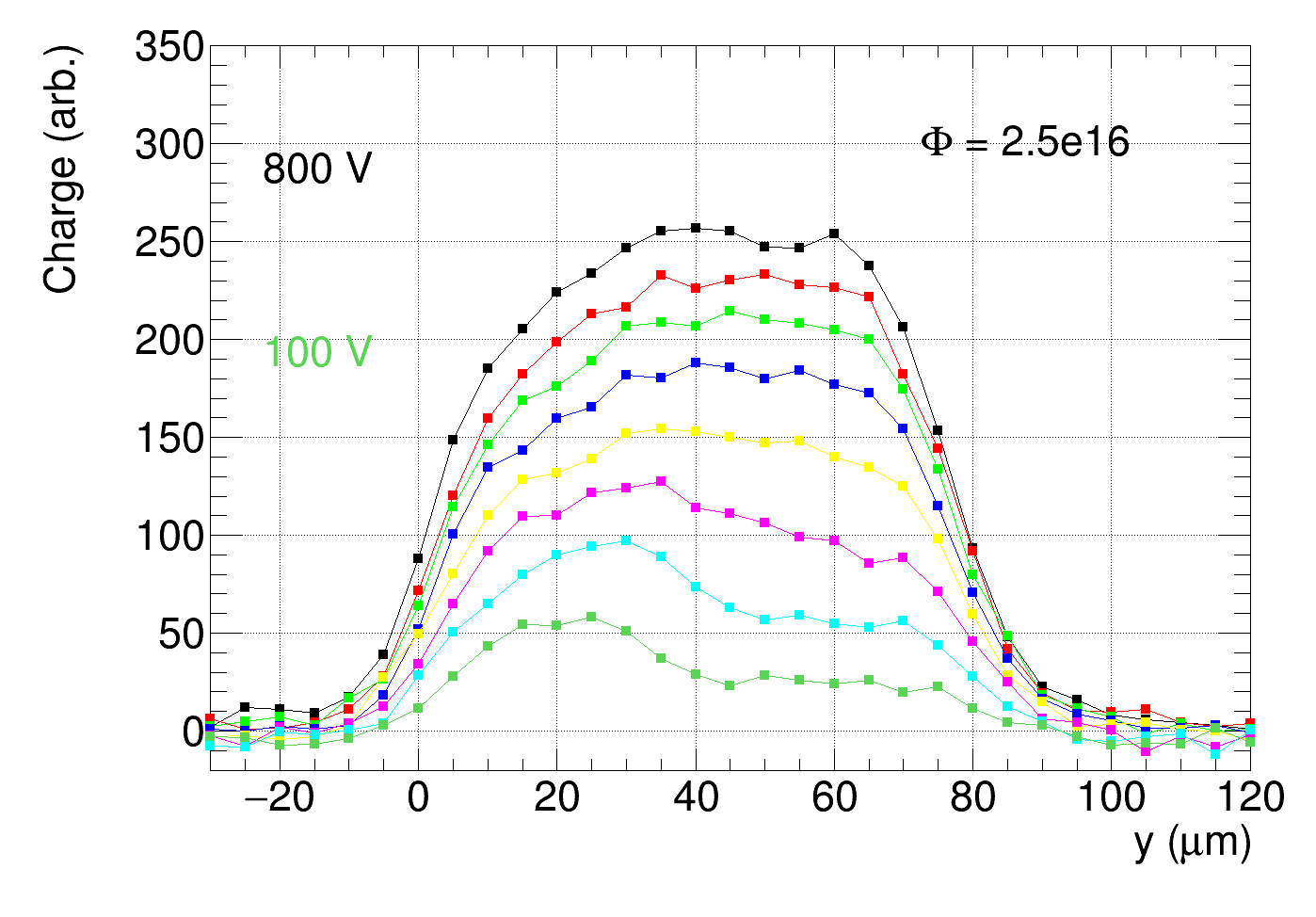} &
\includegraphics[width=.3\textwidth]{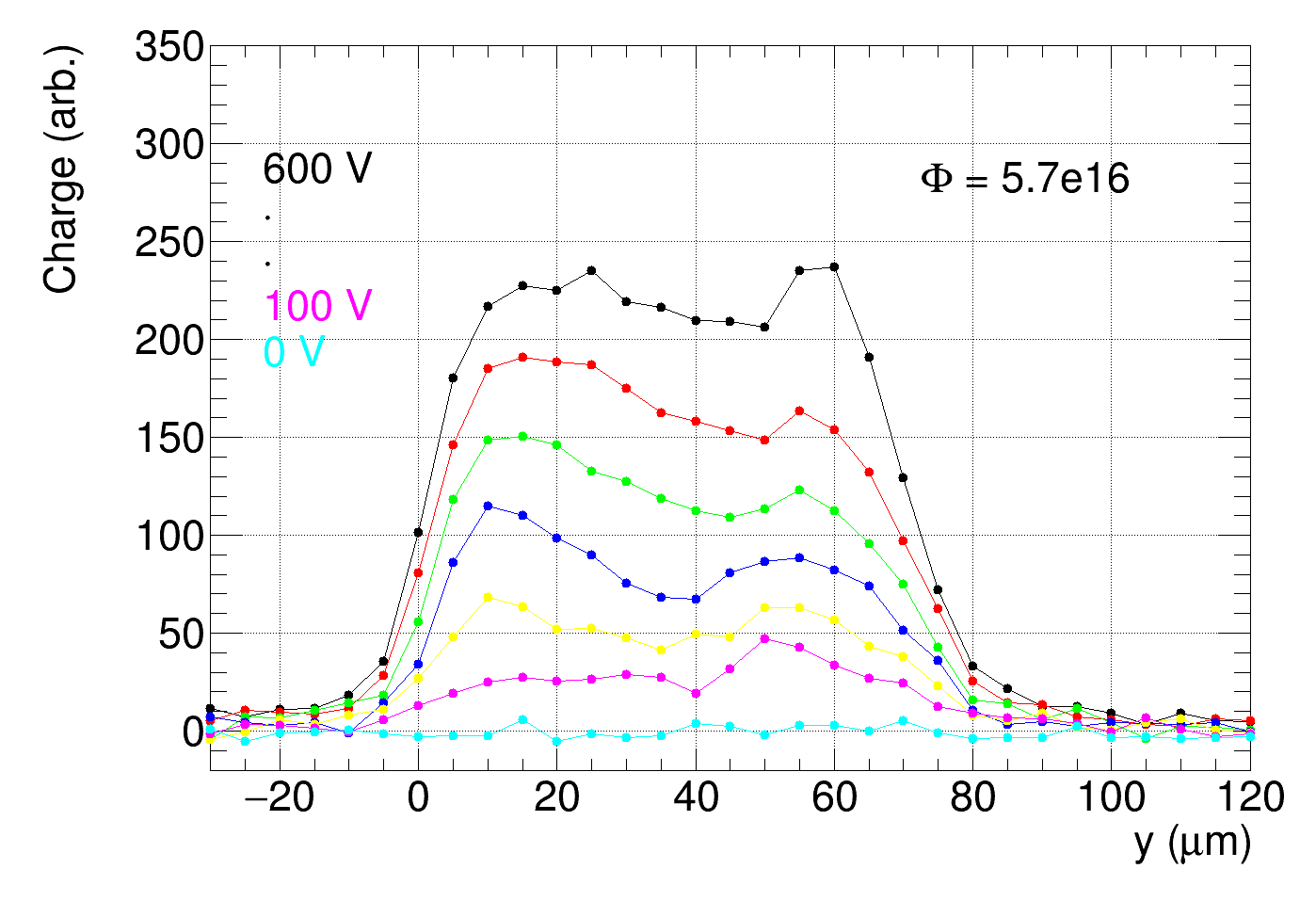} &
\includegraphics[width=.3\textwidth]{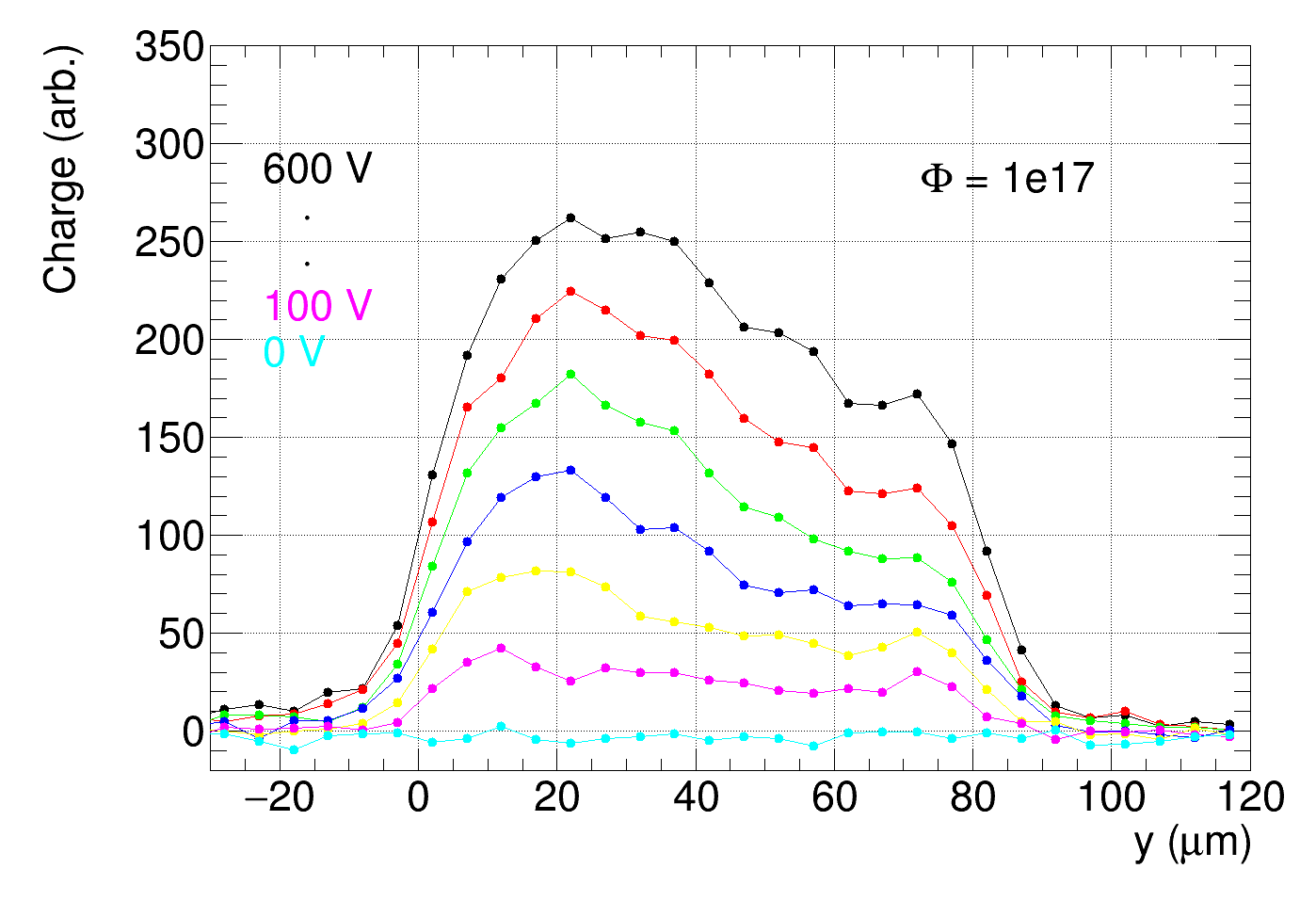} \\
a) & b) & c) \\
\includegraphics[width=.3\textwidth]{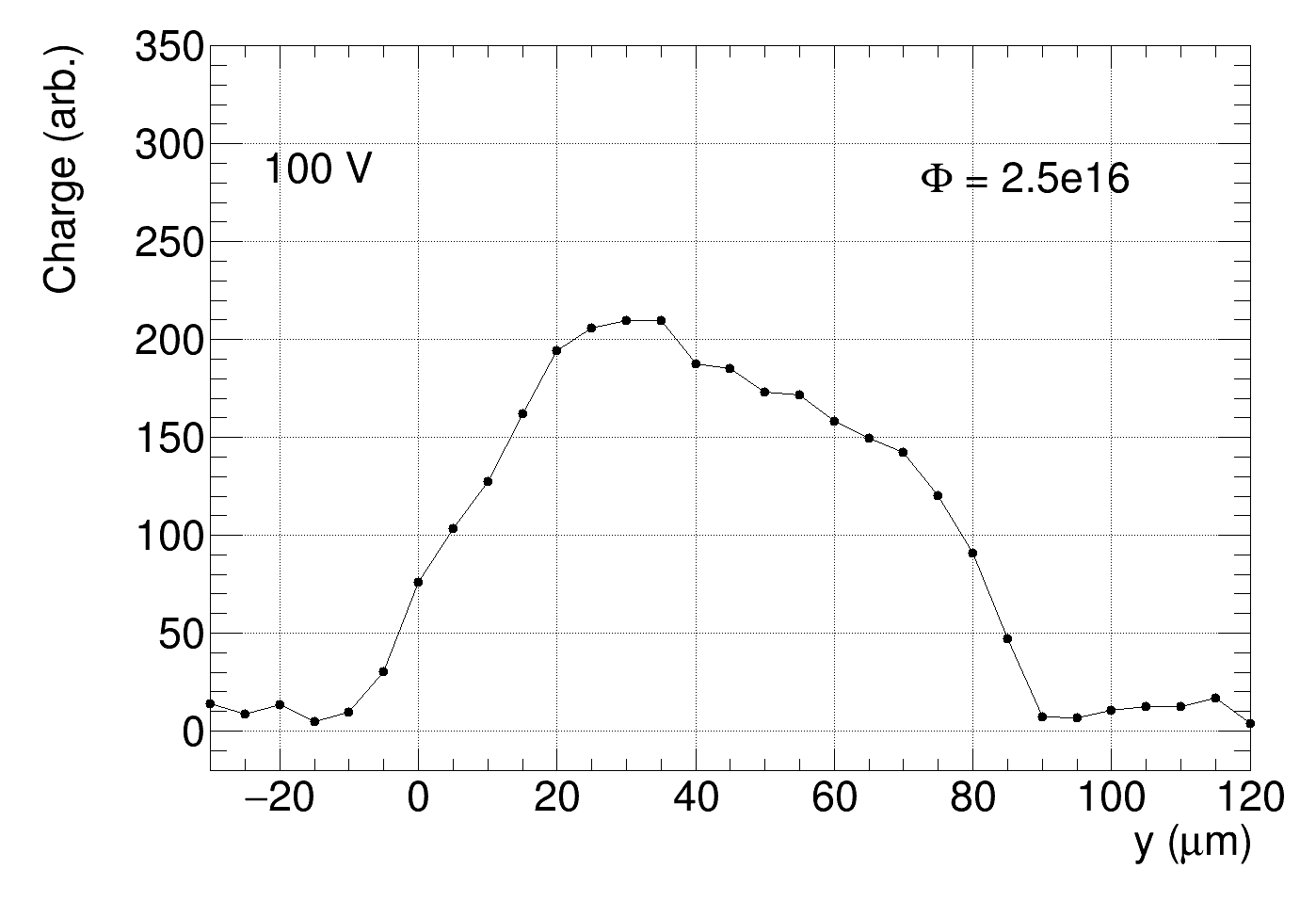} &
\includegraphics[width=.3\textwidth]{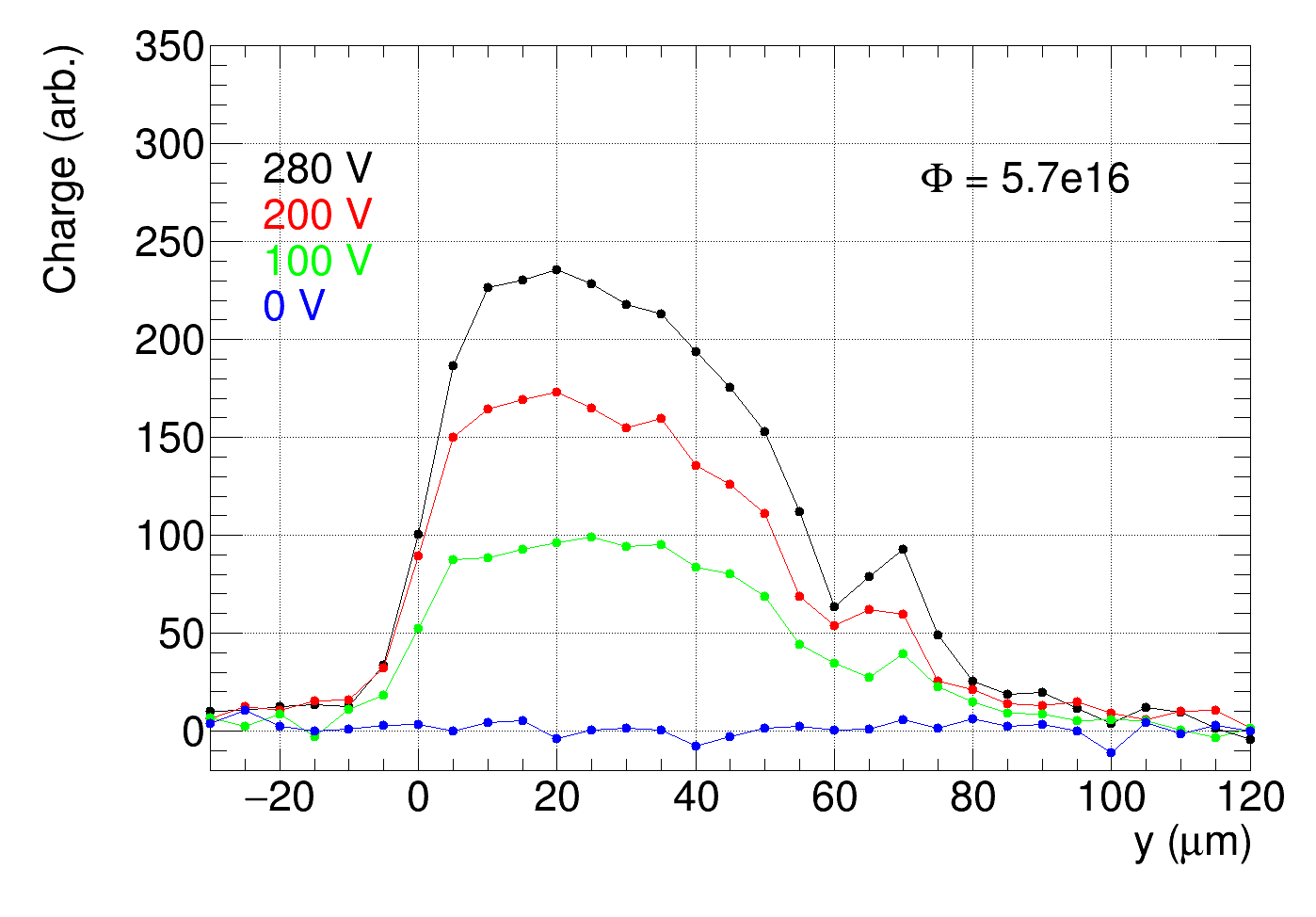} &
\includegraphics[width=.3\textwidth]{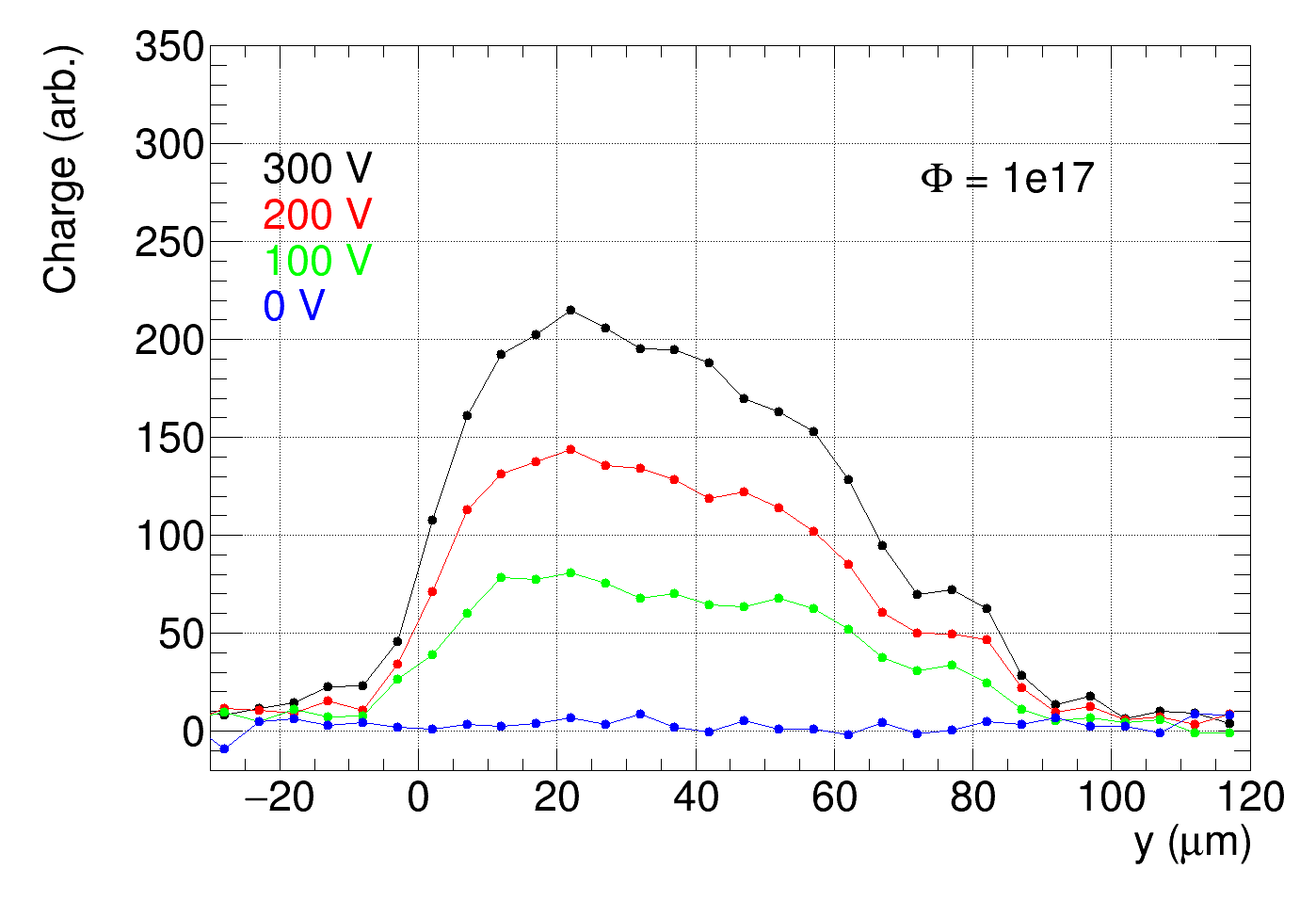} \\
d) & e) & f) \\
\end{tabular}

 \caption{\label{e-tct} Charge collection profiles under reverse bias (a,b,c) and forward bias (d,e,f). Detector surface is at
 $y = 0$ $\mu$m.} 
\end{figure}

As explained in more detail in \cite{Edge-TCT}, in Edge-TCT measurement a short pulse ( < 1 ns) of narrow ($\sim$ 10 $\mu$m) laser beam
($\lambda$ = 1064 nm) is directed to the side of the detector and the current induced on the readout electrode observed using a wide
band amplifier. The output of the amplifier is recorded with a digital oscilloscope. The integral of the recorded waveform is proportional to collected charge. It is calculated offline; an integration time of 10 ns was used in this work. 

Figure \ref{e-tct} shows charge collection profiles i.e. the collected charge as a function of the detector depth for irradiated detectors.
Measurements were made under reverse and forward bias. The highest bias voltage at which measurements were made was limited by
maximal current of 500 $\mu$A. Profiles were measured at $\sim$ -10$^\circ$C which was the lowest temperature achievable with the
system used. It can be seen in Fig. \ref{e-tct} that charge is collected only from the 75 $\mu$m thick epitaxial layer and there is no significant electric field in the low resistivity substrate, including the sample irradiated to highest fluence. Charge seen in the tails of the profile at $y < 0 \mu$m and $y > 75 \mu$m in  Fig. \ref{e-tct} is due to the width of laser beam ($\sim$ 10 $\mu$m) so the depletion depth should be estimated from the distance between 50\% points of rising and falling edges of the charge collection profile.

It can also be noted in Fig. \ref{e-tct} that charge is collected across the entire depth of the epitaxial layer even at lowest reverse bias voltages. This is in agreement with other measurements of heavily irradiated detectors \cite{field,klanner} where significant electric
field is found across the entire bulk and near the back electrode at low bias voltages. This is a consequence of polarization (or double peak) effects in heavily irradiated silicon.
High resistivity of irradiated silicon enables operation at high forward bias voltages
\cite{klanner,fielding,chilingarov} and the electric field in irradiated forward biased detectors is to a large extent a consequence of the high ohmic resistivity of bulk \cite{klanner} material.

\subsection{Charge collection}

Charge collection measurements were made with the experimental setup described in more detail in \cite{ccestup}.
Device Under Test (DUT) is mounted in an aluminium box with 1 mm diameter holes on both sides to collimate electrons from $^{90}$Sr source.
The box is mounted on a cooled plate above a small scintillator, matching approximately the size of the DUT. Only electrons from the high-energy end of the $^{90}$Sr spectrum have sufficient
energy to traverse the silicon detector and trigger the readout by depositing sufficient energy in the scintillator coupled to a photomultiplier. The n$^+$ electrode was connected to readout while guard ring was floating.
When triggered by the photomultiplier a digital oscilloscope records the waveform from the custom made shaping circuit with 25 ns
peaking time connected to the Ortec 142 charge sensitive preamplifier.
The geometry of collimators and sufficient active area of the DUT ensure that more than 97\% of recorded waveforms are associated with the passage
of an electron through the detector. This enables a measurement of collected charge even at very low signal-to-noise ratios when the noise
and the signal peak can not be separated.
In such a case the mean of the signal distribution sampled at the time of the peak of the amplifier-shaper output is
proportional to the mean collected charge in the DUT.
The absolute scale of the system was calibrated by measuring the Most Probable Value (MPV) of collected charge of a standard
300 $\upmu$m thick fully depleted silicon detector and confirmed with 59.5 keV photons from $^{241}$Am source.

Charge collection measurements were made at T = -30$^\circ$C. Detectors were biased with high voltage power supply and the detector current
was monitored during charge collection measurements. Guard ring was left unconnected. Measurements were made under reverse and forward bias.
The bias voltage was increased until large fluctuations of the baseline ($\sim$ 3$\times$ or more larger than at low bias) were observed in the oscilloscope.

\subsubsection{Current}

\begin{figure}[htbp]
  \centering % \begin{center}/\end{center} takes some additional vertical space
 \begin{tabular}{ c c }

     \includegraphics[width=.5\textwidth]{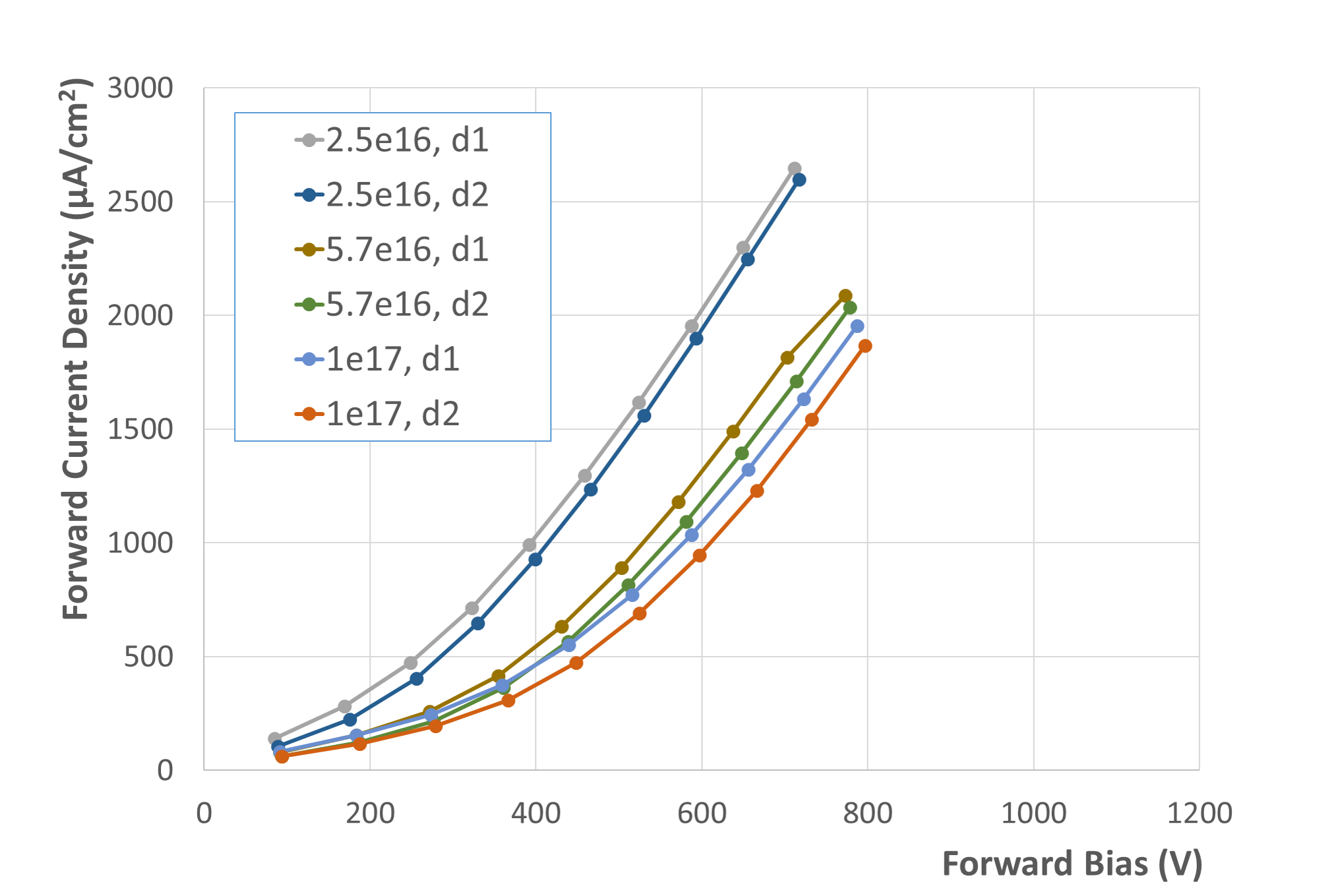} &
     \includegraphics[width=.5\textwidth]{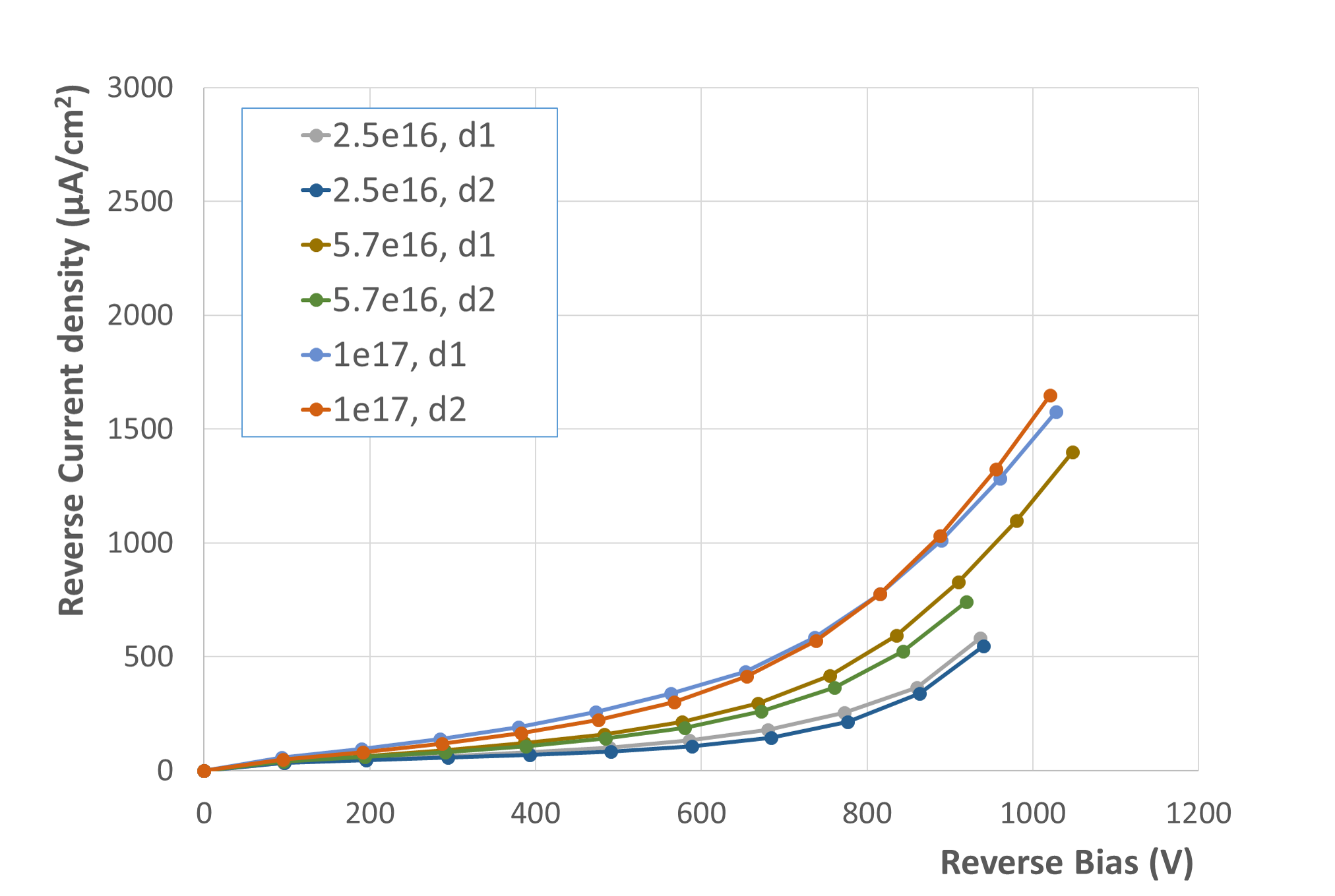} \\
   a) & b) \\
\end{tabular}

 \caption{\label{current} Current density vs. bias voltage at -30$^\circ$C for a) forward bias and b) reverse bias. Measurements with two
   devices per fluence are shown.} 
\end{figure}

Figure \ref{current} shows the detector current density as a function of bias voltage at T = -30$^\circ$C measured in the charge collection setup. The reverse current was measured after annealing for 80 minutes at 60$^\circ$C. Forward current shown in Fig. \ref{current}b was measured after 1200 minutes of annealing at 60$^\circ$C 
becasue forward current measurements at shorter annealing times were not complete. But, as will be shown later, annealing effects on the forward current are small.
Detector current under forward bias is larger than under reverse bias at the same voltage but the difference gets smaller at higher fluences. It can
also be seen that at given voltage under reverse bias the current increases with increasing fluence, while under forward bias it falls, indicating different sources of
current. Under forward bias the current is due to injection of carriers from the junction and it falls with irradiation (at given bias) because of increasing
material resistivity, while under reverse bias carriers also originate from the generation in the depleted bulk.

\subsubsection{Collected charge}

\begin{figure}[htbp]
  \centering % \begin{center}/\end{center} takes some additional vertical space
 \begin{tabular}{ c c }

     \includegraphics[width=.5\textwidth]{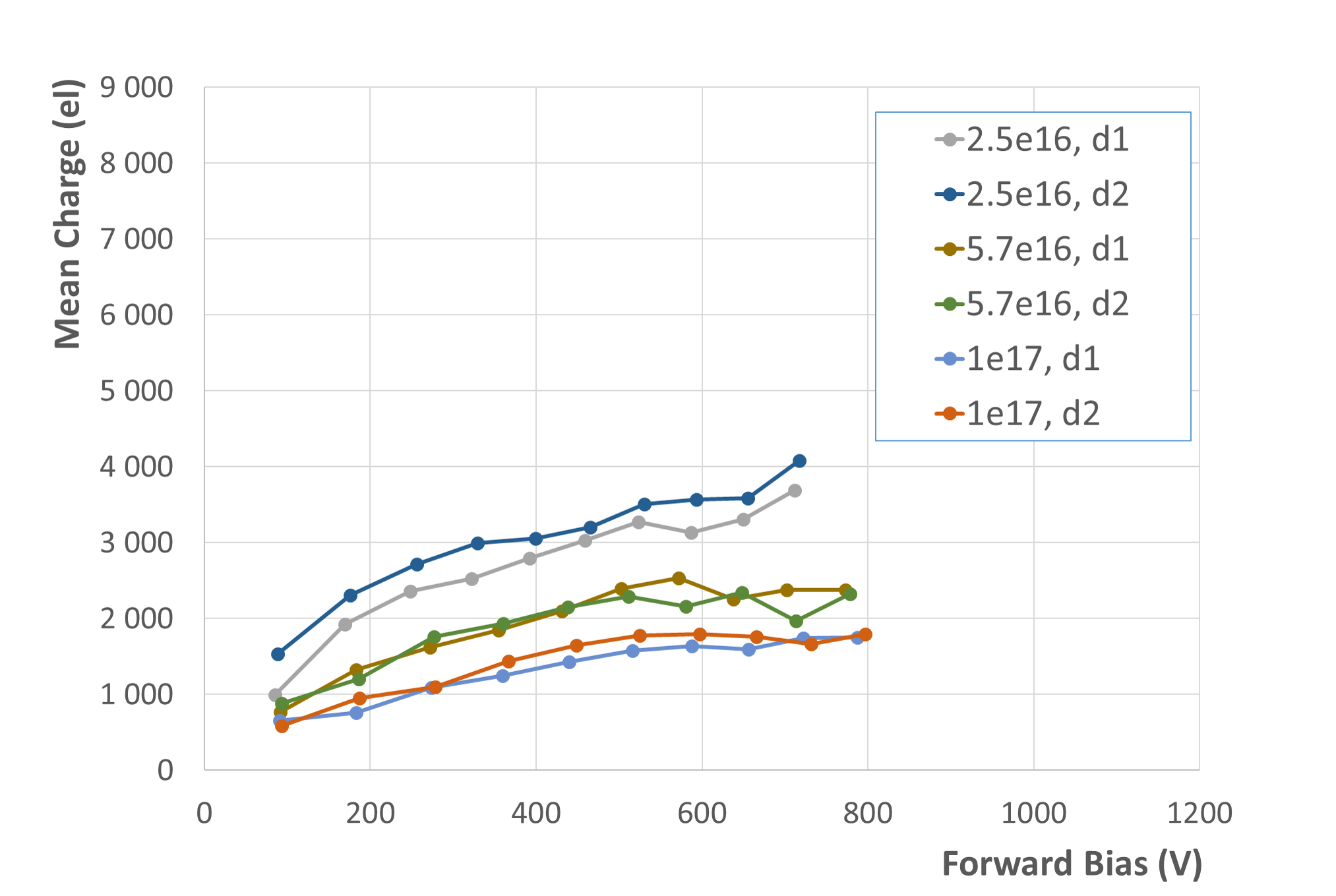} &
     \includegraphics[width=.5\textwidth]{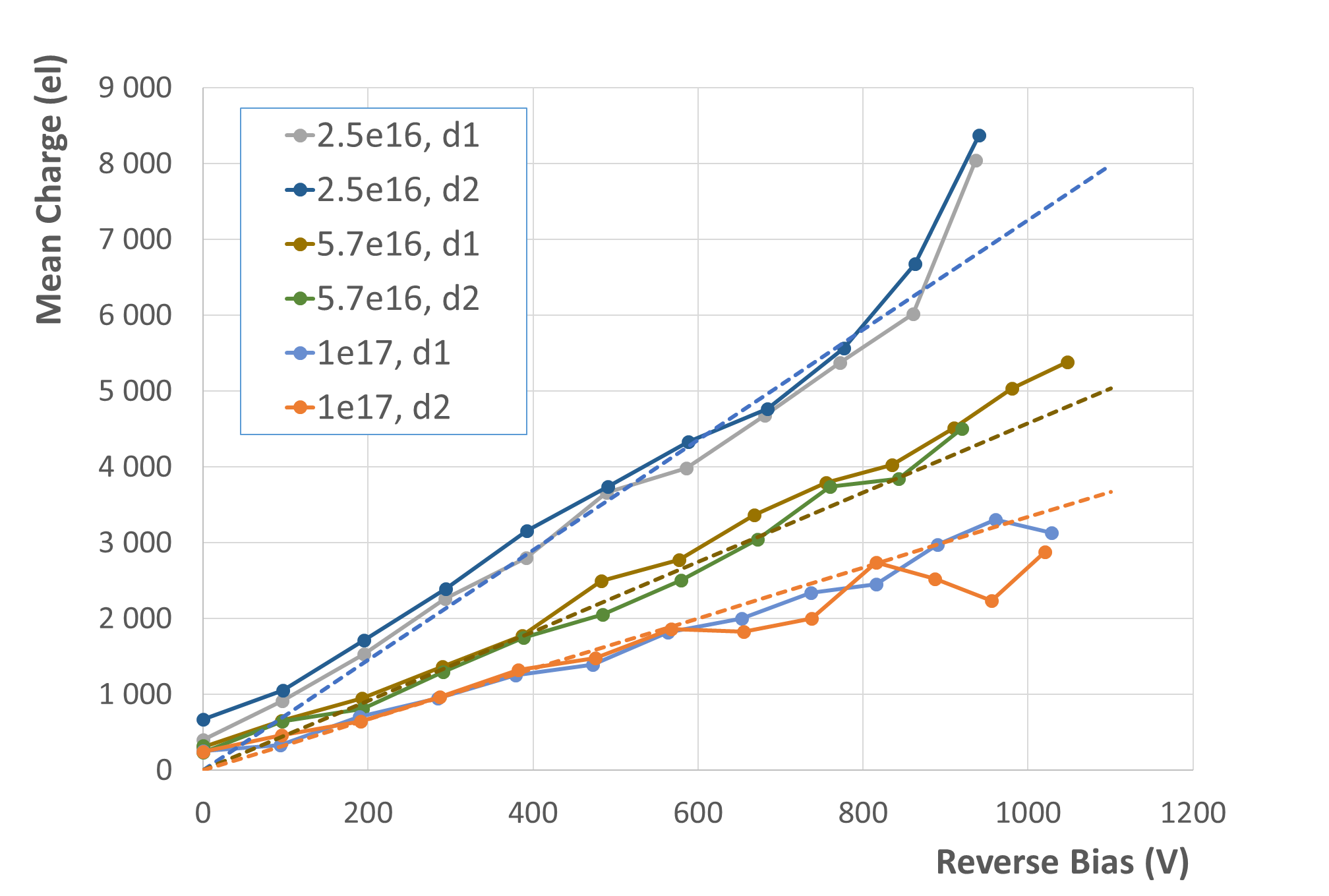} \\
   a) & b) \\
\end{tabular}

 \caption{\label{charge} Mean charge vs. bias voltage for a) forward bias and b) reverse bias at different fluences. Dashed lines in figure b) show the empirical formula \ref{magic} fit to measurements.} 
\end{figure}

Figure \ref{charge} shows the mean charge as a function of forward and reverse bias voltage. Mean charge is shown because in these measurements signal/noise ratio is very low and spectra can be well fitted with a Gaussian function. At bias voltages below $\sim$ 400 V more charge is measured under
forward bias while at higher voltages charge collection under reverse bias is larger.

In reverse bias the 
mean charge increases almost linearly with increasing voltage. Somewhat faster increase can be seen above 800 volts
for the lowest fluence which is an indication of charge multiplication. Linear increase with the reverse bias voltage at extreme fluences was
measured also in \cite{spagheti} and an empirical formula describing scaling of the collected charge with fluence and bias voltage
was introduced. The same ansatz was also explored here:
\begin{equation}
  Q = k\cdot \Phi^{b}\cdot V,
  \label{magic}
\end{equation}
where $Q$ is the mean collected charge, $\Phi$ the equivalent fluence in units of $10^{15}$ n/cm$^2$ and $V$ the reverse bias voltage. Parameters $k = 44$ el/V and $b = -0.56$ were determined from the fit to the measured points. The formula \ref{magic} has no direct physical explanation but it is useful for a rough estimation of the collected charge.
In \cite{spagheti} parameter values $k_{300} = 26.4$ el/V and $b_{300} = -0.683$ were found for most probable charge in 300 $\mu$m thick detectors.
Evaluating equation \ref{magic} for $V$ = 600 V and $\Phi = 10^{17}$ n/cm$^2$ yields $Q$ = 2000 el and for 300 $\mu m$ detector with $k_{300}$ and $b_{300}$ one gets $Q_{300,MPV}$ = 680 el which should be multiplied with 1.25 $Q_{300}$ = 850 el to take into account that equation \ref{magic} describes the mean charge and not the most probable value measured with 75 $\mu$m thick detector.  This clearly shows that at extreme fluences significantly more charge is collected with thin detectors at same reverse bias voltage. 

\begin{figure}[htbp]
  \centering % \begin{center}/\end{center} takes some additional vertical space
 \begin{tabular}{ c c }

     \includegraphics[width=.5\textwidth]{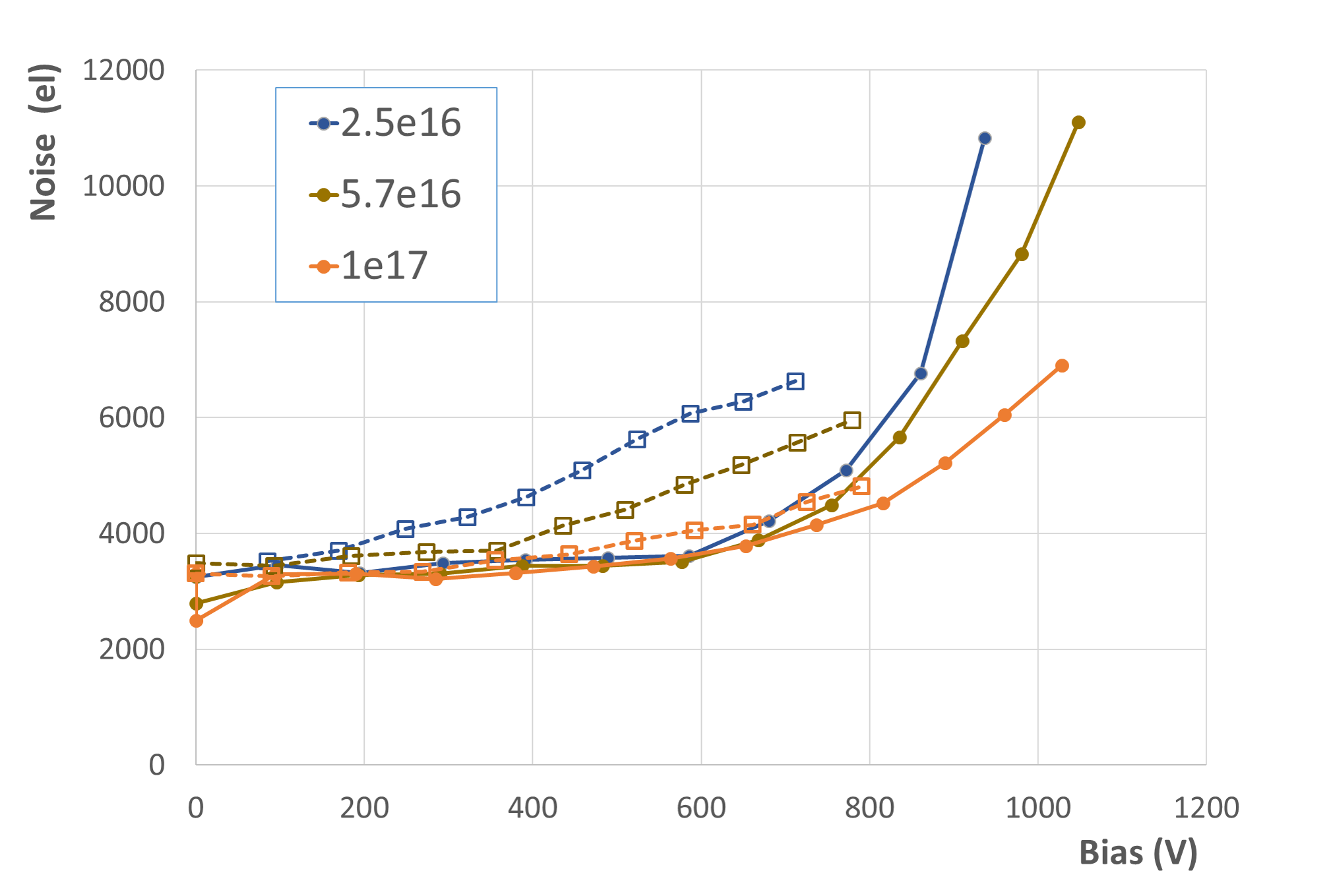} &
     \includegraphics[width=.5\textwidth]{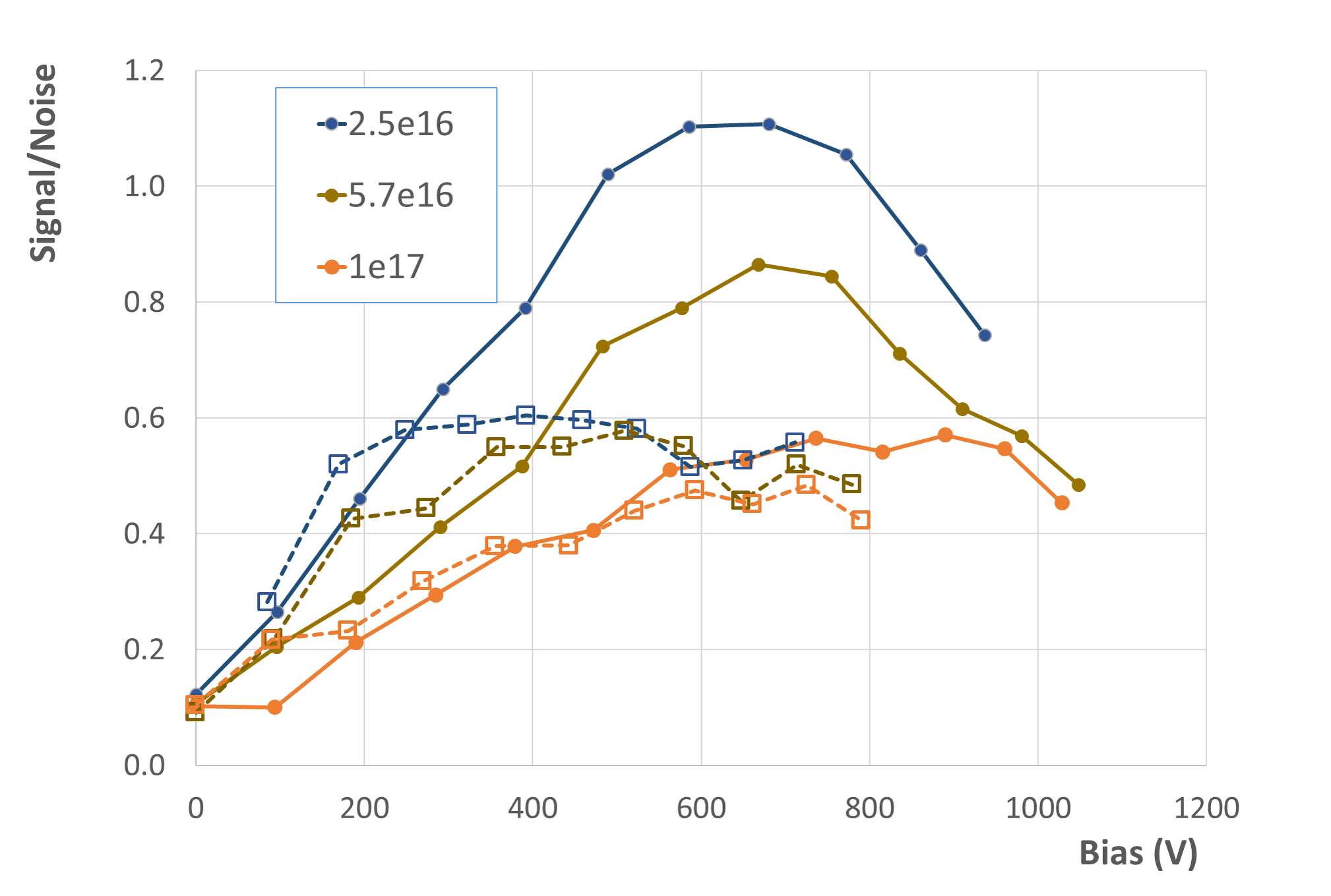} \\
   a) & b) \\
\end{tabular}

 \caption{\label{noise} Noise a) and ratio Signal/Noise b) vs. voltage for forward and reverse bias. Full symbols are for reverse bias and empty symbols for forward bias, fluence is color coded. Values for one sample per fluence is shown.} 
\end{figure}

Increase of bias voltage is accompanied with an increase of detector current leading to increased noise thus affecting the
signal/noise ratio. At high bias voltages fluctuations are enhanced because of charge multiplication and breakdown effects.

In addition, approaching operation with gain and breakdown increases fluctuations of collected charge even further. 
Figure \ref{noise} shows noise and signal/noise ratio as a function of bias voltage. Noise is measured from the distribution of
outputs sampled from the baseline well before the arrival of the trigger. The distribution is fit by the Gaussian function and noise is
taken as the standard deviation $\sigma$ obtained from the fit. Signal used to calculate the signal/noise ratio in Fig. \ref{noise}b) is defined as
the mean charge shown in Fig. \ref{charge}.
It can be seen in Fig. \ref{noise}a) that at lower bias voltages noise increases slowly because of the increase of detector current. The increase of noise
is larger in forward bias but differences are smaller at higher fluences following the behaviour of detector current. Above $\sim$ 700 V the reverse bias
noise starts to increase faster reflecting the onset of multiplication. Signal/noise ratio in \ref{noise}b) shows that the best charge collection
performance would be achieved at the reverse bias of $\sim$ 600 V. The benefit of higher charge measured at higher voltage (Fig. \ref{charge}) is lost by
the large increase of noise. It can also be seen that at the highest fluence there is almost no difference between forward and reverse biases.

It is important to stress that both the type of detector and the measurement system are not optimized for low noise operation and results in Fig. \ref{noise} should be understood as an example of the possible trend of the S/N in highly irradiated
detectors. Signal/noise performance of detector with smaller pixel sizes and different breakdown behaviour might be different
but similar effects would govern the optimal detector bias choice.  

\section{Annealing}

Charge collection measurements were repeated after several steps of annealing at 60$^\circ$C: after 80, 240, 560 and 1200 minutes.

\begin{figure}[htbp]
  \centering % \begin{center}/\end{center} takes some additional vertical space
 \begin{tabular}{ c c }

     \includegraphics[width=.5\textwidth]{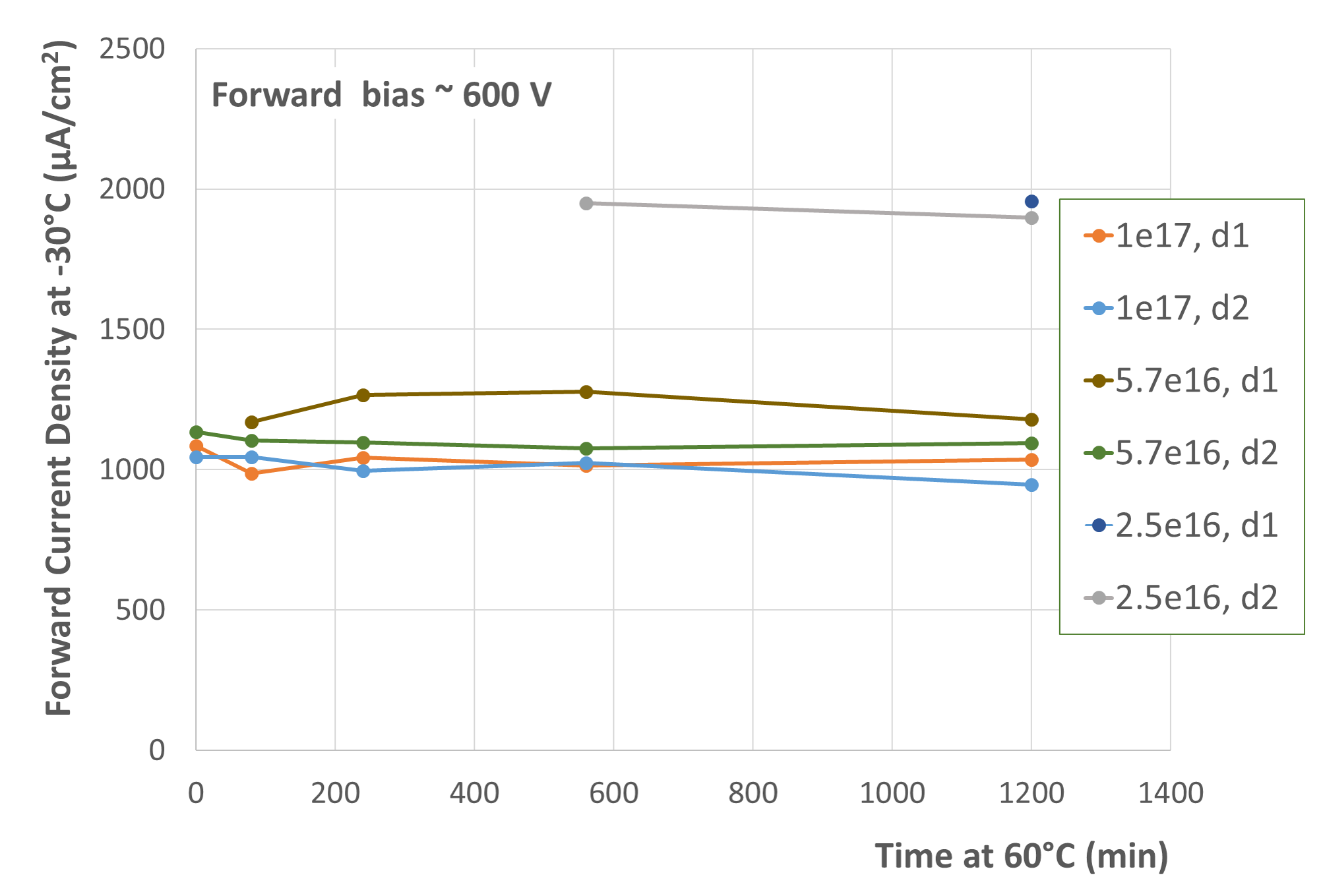} &
     \includegraphics[width=.5\textwidth]{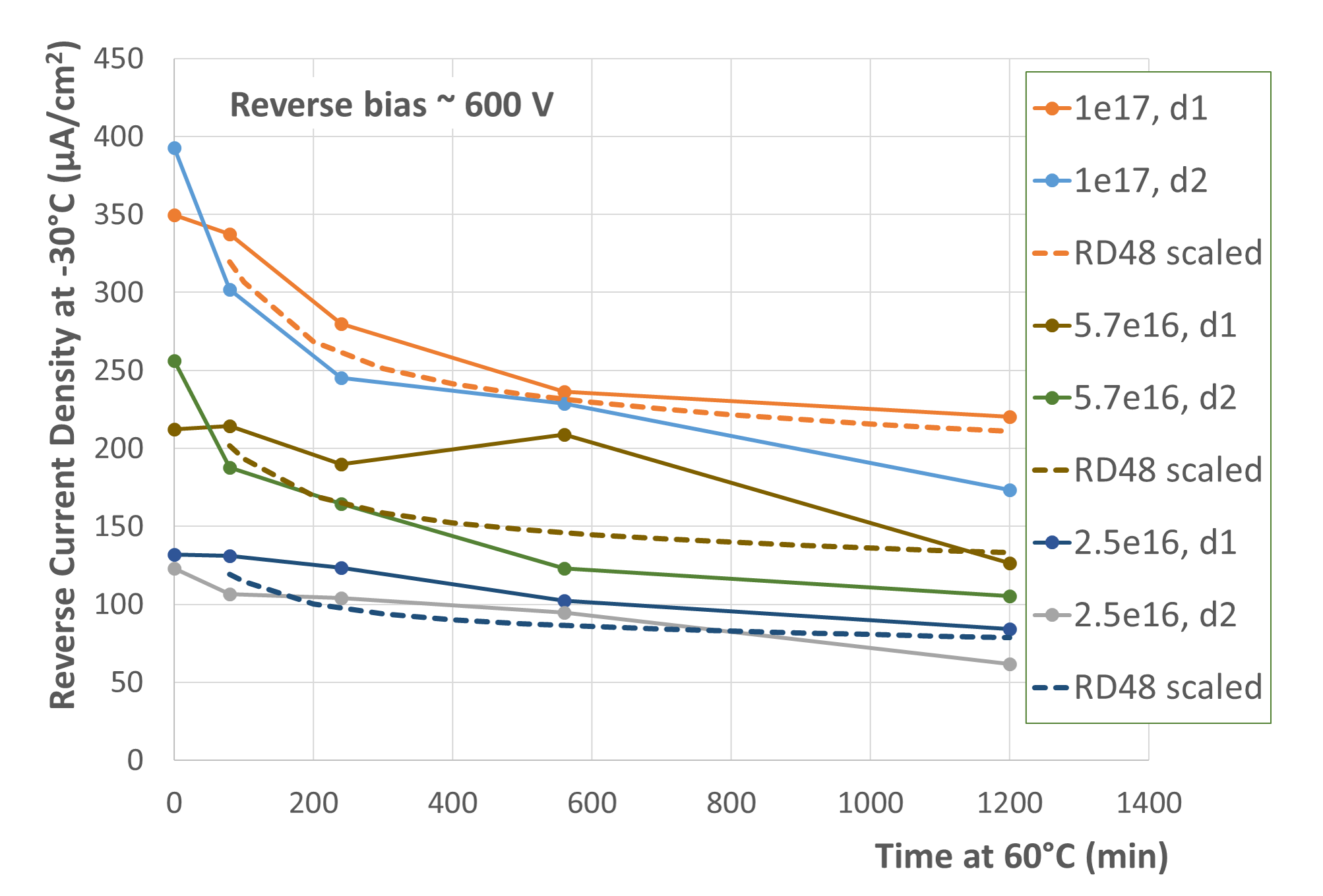} \\
   a) & b) \\
\end{tabular}

 \caption{\label{curraneal} Forward a) and reverse b) current density measured at 600 V as a function of accumulated annealing time at 60$^\circ$C. Dashed lines in figure b) show calculated annealing curves as explained in the text.} 
\end{figure}

Figure \ref{curraneal} shows forward and reverse current densities measured at bias of 600 V as a function of annealing time. Measurements under forward bias were not
made at all annealing times for all fluences. Results in Fig. \ref{curraneal}a) indicate that forward bias current at a given bias voltage doesn't change significantly
with annealing time. Under reverse bias the detector currents decrease with annealing time, as shown by Fig. \ref{curraneal}b). 
The dashed lines in figure \ref{curraneal}b) show the calculated time dependence of the annealing of the leakage current at 60$^\circ$C according to parametrisation
developed by the RD48 collaboration \cite{moll}: $I \propto \alpha_1 \exp(-\frac{t}{t_1}) + (\alpha_0-\beta \ln{t})$ with
parameters $\alpha_1 = 1\times 10^{-17}$ Acm$^{-1}$, $t_1$ = 93 min, $\alpha_0 = 5\times 10^{-17}$ Acm$^{-1}$ and $\beta = 3.3\times 10^{-18}$ Acm$^{-1}$ and
time $t$ in minutes. The proportionality factor was determined so that calculated values match the average of the two measurements at 80 minutes for each fluence.
It can be seen in \ref{curraneal}b) that the calculated time dependence roughly follows the measured points.

It should be mentioned that RD48 function describes annealing of the reverse current in a fully depleted silicon detector where current generation volume is well defined.
In this work current is increasing with bias voltage, as can be seen in Fig. \ref{current}b, and it is not possible to determine the voltage of full depletion.
Bias of 600 V was chosen because this is the highest voltage before current (Fig. \ref{current}b) and noise (Fig. \ref{noise}b) start to increase 
more rapidly. Measured currents after 80 minutes annealing were by about factor of 2 to 3 higher than calculated even if it was assumed that current was generated in the whole detector volume (3 mm$^2$ $\times'$ 75 $\mu$m). This indicates that multiplication effects might significantly contribute to the current at these fluences.

\begin{figure}[htbp]
  \centering % \begin{center}/\end{center} takes some additional vertical space
 \begin{tabular}{ c c }

     \includegraphics[width=.5\textwidth]{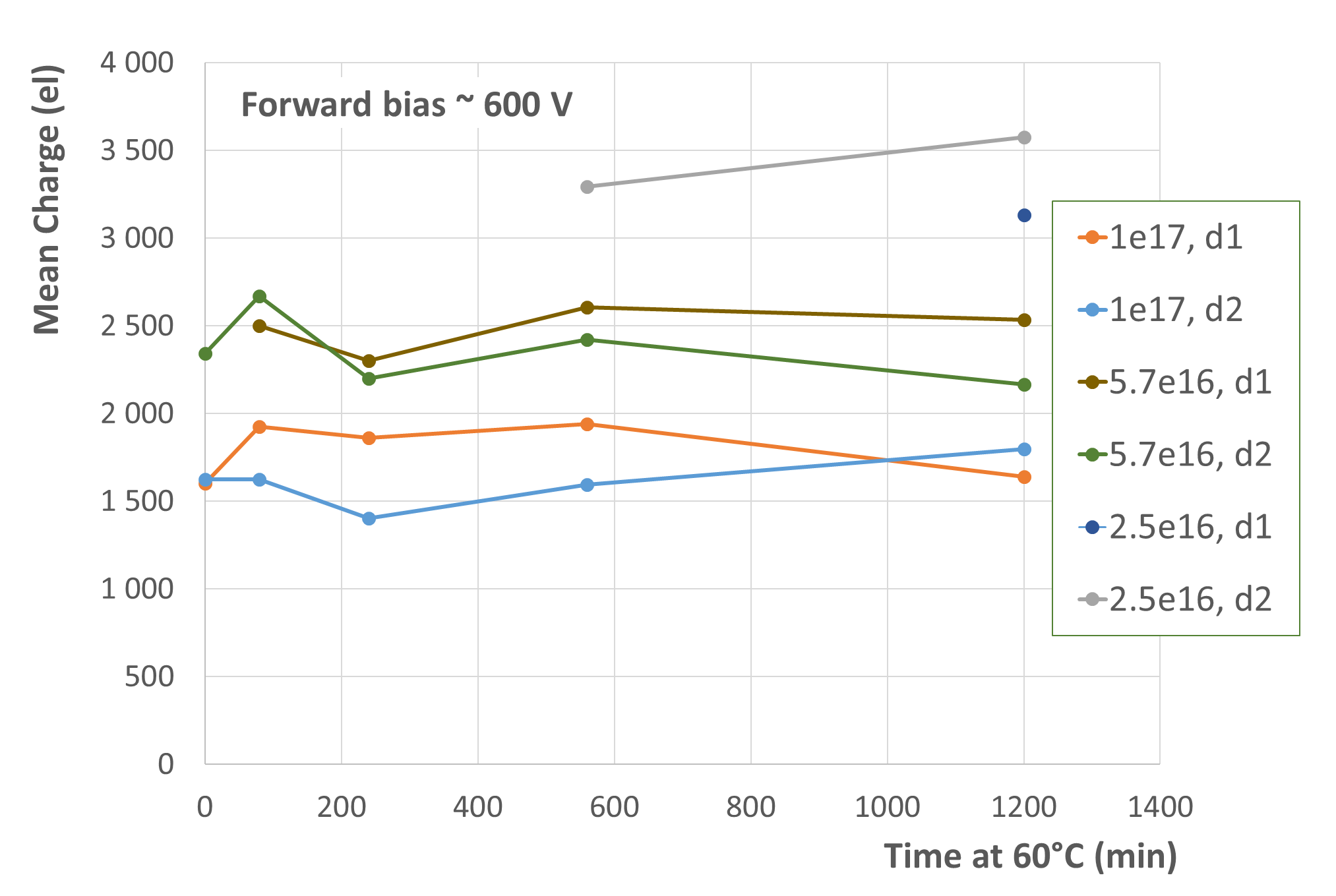} &
     \includegraphics[width=.5\textwidth]{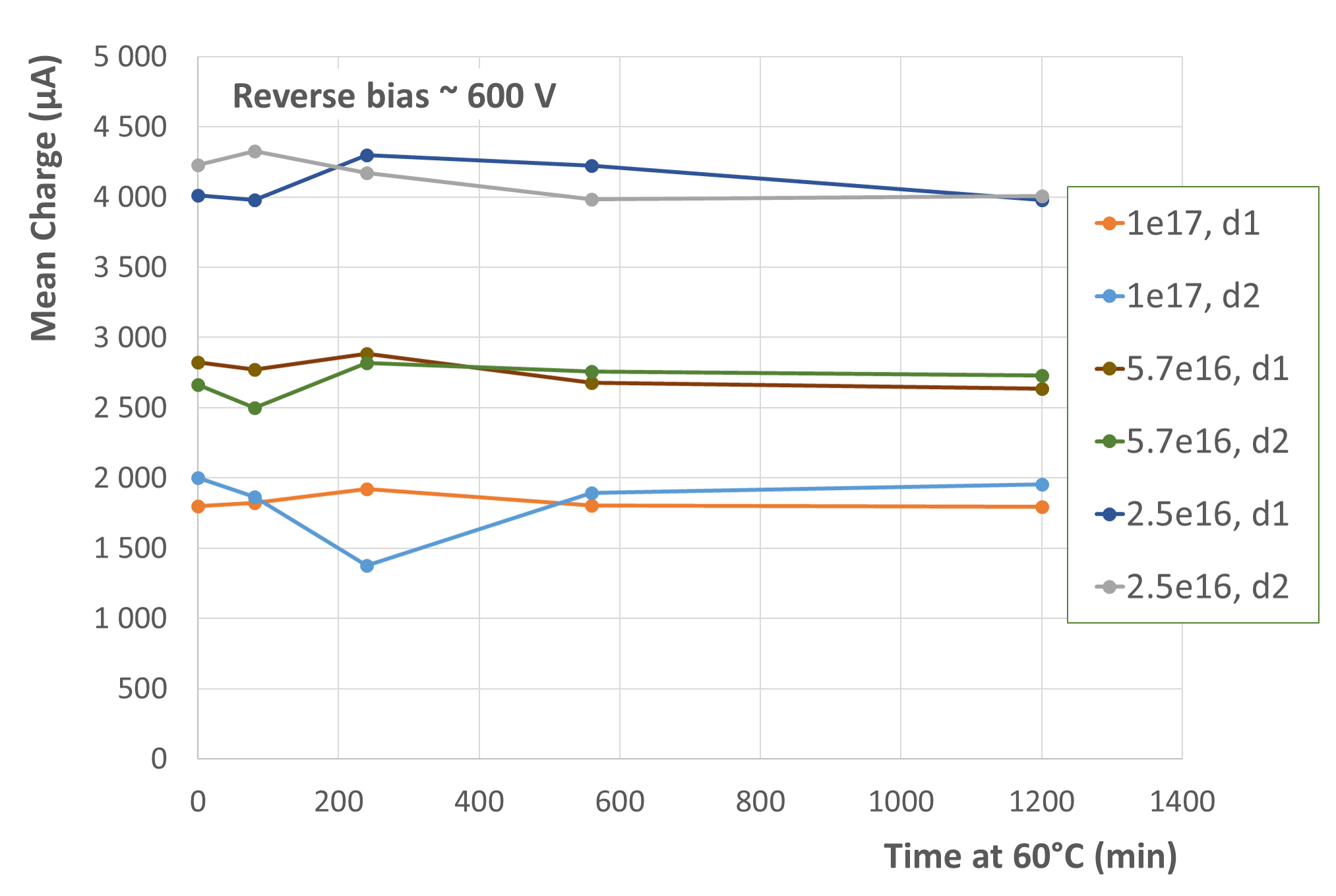} \\
   a) & b) \\
\end{tabular}

 \caption{\label{chargeanneal} Collected charge at 600 V for forward a) and reverse b) bias as a function of accumulated annealing time at 60$^\circ$C.} 
\end{figure}

Figure \ref{chargeanneal} shows the collected charge as a function of annealing at 60$^\circ$C. No significant effect can be seen for  forward or reverse bias at 600 V. No significant effect of annealing was also observed at different bias voltages.

\section{Effective Trapping Times}

Measurements of collected charge vs. forward bias voltage offer the opportunity to estimate trapping times at extreme fluences.
This is especially important as measurements indicate that linear scaling of effective trapping probability with fluence
$1/\tau = \beta \cdot \Phi_{eq}$, measured at $\mathcal{O}$ (10$^{14}$ n/cm$^2)$ with $\beta \sim 5\times 10^{-16}$ cm$^2$ns$^{-1}$
\cite{trap-krambi,trap-cindro}, overestimates measured trapping times by 50\% already at fluences of 3 $\times 10^{15}$ n$_{eq}$/cm$^2$  \cite{trap-CMS}. Trapping time is one of the basic parameters needed for modeling of detector performance.

Charge collection measurements in forward bias are suitable for estimation of trapping time because it can be assumed that the
electric field in a pad detector is not far from constant $E = V/D$ where $V$ is the forward bias voltage and $D$ the detector thickness.
At high fluences trapping times are short and drift distances of charge carriers are shorter than the detector thickness. At fluence $\Phi_{eq} = 10^{16}$ n/cm$^2$ the effective trapping time can be estimated as $\tau$ = 1/($\Phi \cdot \beta$) = 0.2 ns which at saturated velocity $v_{sat} \sim $100 $\mu$m/ns yields a drift distance of 20 $\mu$m. This is significantly shorter
than the 75 $\mu$m detector thickness even if trapping times are underestimated by a factor of 2. In this case it can be roughly approximated that collected charge $Q$ in pad detector is:
\begin{equation}
  Q = \frac{\Delta Q}{\delta x}\cdot D \cdot \frac{1}{D} \cdot v \cdot \tau = \frac{\Delta Q}{\delta x} \cdot v \cdot \tau,
  \label{colcharge}
\end{equation}
where $\frac{\Delta Q}{\delta x}$ describes the mean charge per unit track length released by a MIP, $D$ is the detector thickness which cancels with the $\frac{1}{D}$ factor describing the weighting field. $v = v_e + v_h$ is the sum of electron and hole velocities and $\tau$ is the trapping time. The same value of trapping time is taken for electrons and holes.
Expression \ref{colcharge} is accurate to better than 25\% if the carrier drift distance is shorter than $D/2$ \cite{irena_doc}.

Carrier velocities depend on the electric field and also on fluence because of scattering on ionized defects introduced by irradiation. This dependence was estimated up to extreme fluences from Edge-TCT \cite{mikuz-trento} measurements with forward biased detectors.
Induced current measured with Edge-TCT immediately after the laser pulse is proportional to the sum of carrier velocities \cite{Edge-TCT} at the location of laser beam.
As already mentioned above, in forward biased heavily irradiated detectors the electric field can be approximated with a constant $E = V/D$. The dependence of sum of carrier velocity $v$ on electric field $E$ was in \cite{mikuz-trento} parametrized with:
\begin{equation}
v = v_e + v_h = \frac{\mu_{0,\mathrm{sum}} E}{1+\frac{\mu_{0,\mathrm{sum}} E}{v_{e,sat}+v_{h,sat}}}
\label{v_on_e}
\end{equation}
where $v_{e,sat}+v_{h,sat}$ is the sum of saturation velocities for electrons and
holes respectively. Relation \ref{v_on_e} assures that at high bias voltages the saturation velocity is reached 
and the only assumption is that saturation velocities don't depend on fluence. Saturation velocity is determined by optical phonon emission \cite{jacoboni} whose energy depends on properties of the basic Si crystal cell which should not be influenced by irradiation.

Parameter $\mu_{0,\mathrm{sum}}$ represents an effective ``sum'' zero field mobility.
It has been estimated in \cite{mikuz-trento} by fitting equation \ref{v_on_e} to measured values of $v$ as
a function of forward bias. The measurements were performed with Edge-TCT at -20$^{\circ}$C with sensors irradiated with neutrons to fluences ranging from  $\Phi_{eq} = 5\cdot 10^{15}$ n/cm$^2$ to $\Phi_{eq} = 1\cdot 10^{17}$ n/cm$^2$. An empirical scaling of parameter $\mu_{0,\mathrm{sum}}$ with fluence was proposed in \cite{mikuz-trento}:

\begin{equation}
  \mu_{0,\mathrm{sum}} = 3500 \, \mathrm{cm}^2/\mathrm{Vs} \cdot (\frac{\Phi}{10^{15} \mathrm{n/cm}^2})^{-0.46}, 
  \label{muscal}
\end{equation}
which can be used for $\Phi > 5\times 10^{15}$ n/cm$^2$.

From \ref{muscal} it can be estimated that already at $\Phi \sim 5\times 10^{15}$ n/cm$^2$ the mobility is significantly lower than in unirradiated silicon. This can be explained by scattering on ionized defects introduced by irradiation.

\begin{figure}[htbp]
  \centering % \begin{center}/\end{center} takes some additional vertical space
 \begin{tabular}{ c c }

     \includegraphics[width=.5\textwidth]{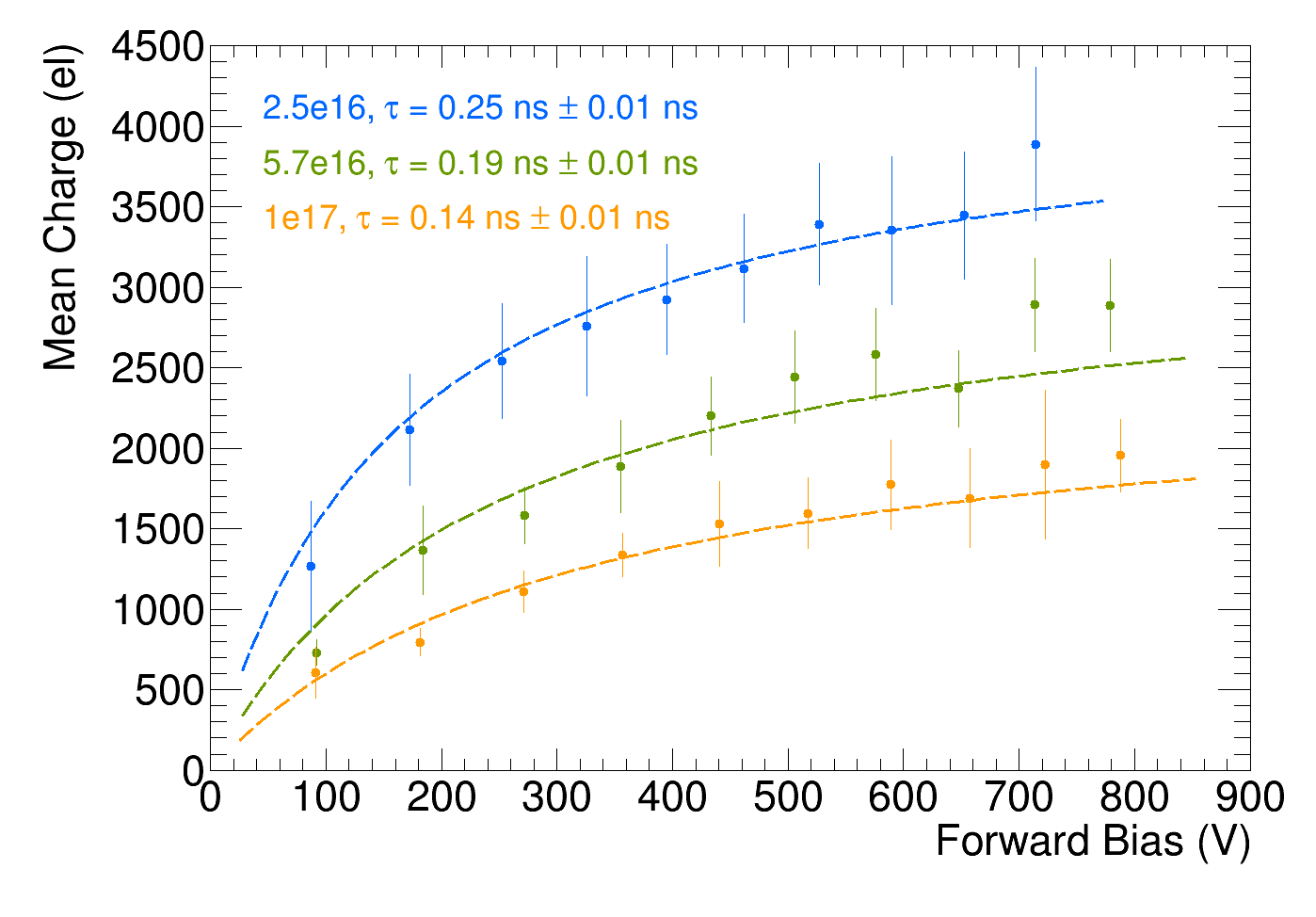} &
     \includegraphics[width=.5\textwidth]{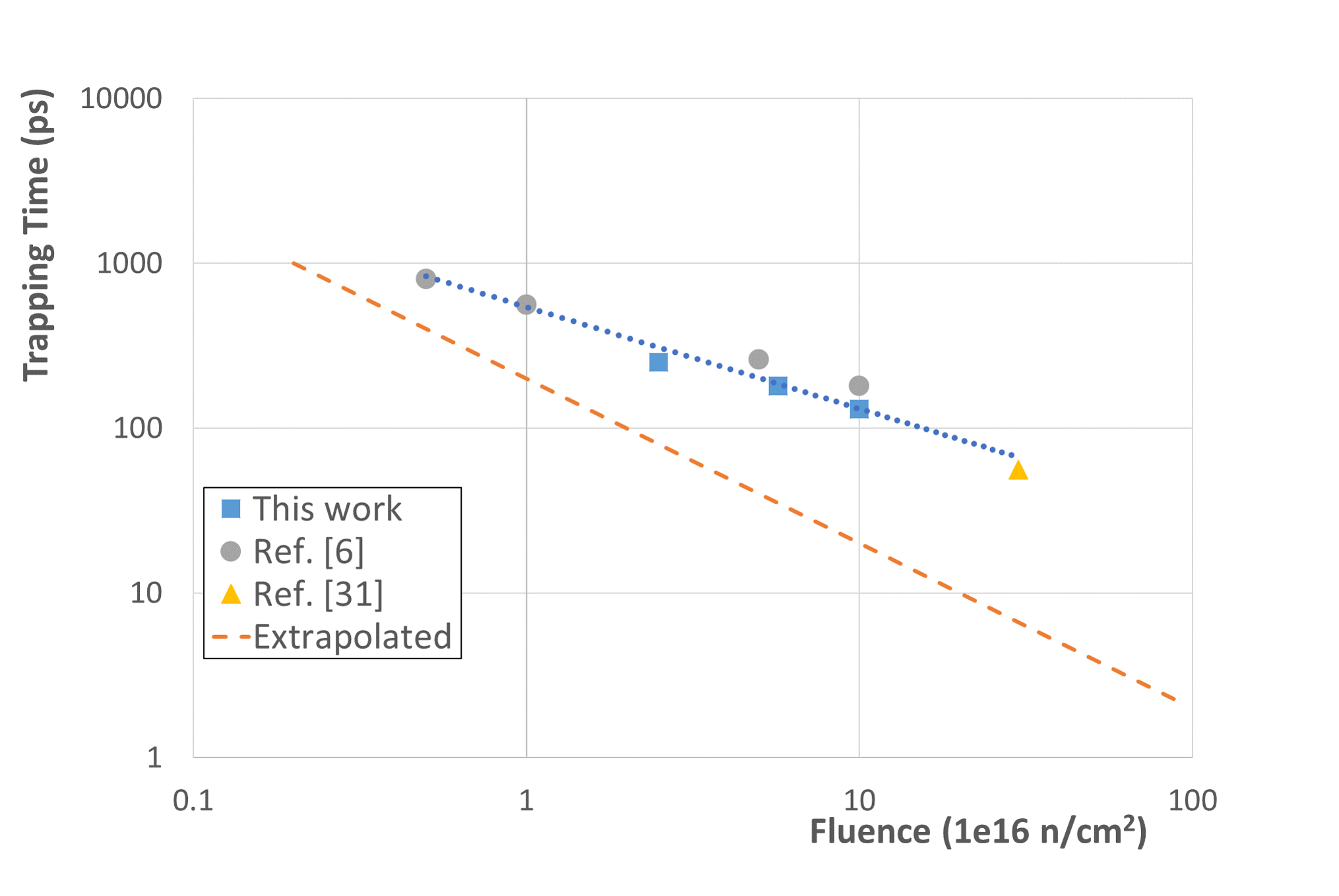} \\
   a) & b) \\
\end{tabular}
 \caption{\label{fitcharge} a) Charge as a function of forward bias fitted with \ref{colcharge} as explained in the text for different fluences. Averages of measurements with two devices at each fluence are shown. Figure b) shows comparison of trapping times from the fit with other measurements and with extrapolation with $\beta = 5\times 10^{-16}$ cm$^2$ns$^{-1}$. Dotted line is a power function fitted to measurements.}
\end{figure}

From dependence of drift velocity $v$ on electric field and fluence measured with Edge-TCT, trapping times can be estimated from the fit of Eq. \ref{colcharge} to
the collected charge measured as a function of forward bias voltage. For the fit the sum of saturation velocities at
T = -30$^{\circ}$C was set to $v_{e,sat}+v_{h,sat} = 175 \mu$m/ns \cite{vsat} and sum zero field mobility calculated
from \ref{muscal} was also scaled to
 T = -30$^{\circ}$C. Electric field was calculated as $E = V/D$.     
 Figure \ref{fitcharge}a) shows the result of the fit of Eq. \ref{colcharge} to measured collected charge at different bias voltages. Measured points in Fig. \ref{fitcharge}a) are averages from two devices per fluence shown in Fig. \ref{charge}a and error bars are RMS of measurements and 10\% systematic error added in quadrature. It can be seen that Eq. \ref{colcharge} fits the measurements well. It is different from almost linear behaviour in reverse bias shown in Fig. \ref{charge}b).
Linear dependence under reverse bias indicates that the electic field is not uniform and that at higher bias voltages multiplication effects contribute significantly. 
 
The trapping times extracted from the fits of Fig. \ref{fitcharge}a) are much longer than values extrapolated using $\tau = 1/(\beta\cdot\Phi)$ with $\beta \sim 5\times 10^{-16}$ cm$^2$ns$^{-1}$.
This is clearly shown in Fig. \ref{fitcharge}b) where the trapping times estimated with the above method and other literature \cite{mikuz-trento,jaztrento} results are compared with trapping
times calculated at these high fluences using parameters measured at lower ones.

Decrease of measured trapping times with fluence is less steep and it follows power function $\tau [\mathrm{ps}] = 54\cdot \Phi^{-0.62}$, where $\Phi$ is equivalent fluence in 10$^{16}$ n$_{eq}$/cm$^2$. Difference between measured and extrapolated values increases with increasing fluence and it is approaching one order of magnitude at 10$^{17}$ n/cm$^2$.
Direct comparison of effective trapping times estimated under reverse and forward bias may be questioned because of significantly different detector currents in these two modes of operation which might influence the occupation of localized energy levels contributing to trapping. However it was shown in \cite{DClighttrap} that even larger variation of current did not lead to significant changes of effective trapping times.   

\section{Conclusions}

In this paper measurements with thin pad detectors irradiated to extreme fluences in the range from  2.5$\times 10^{16}$ up to 1$\times 10^{17}$ n/cm$^2$ are described. Measurements were made with 75 $\mu$m thick p-type epitaxial Si detectors made on low resistivity substrate providing mechanical support and electrical contact with metallized back plane. Edge-TCT
measurements were made to check the active thickness of detectors and lso confirmed that at highest fluences no significant electric field was found in the low resistivity substrate.

Detector current was measured as a function of bias voltage in reverse and forward bias. It was shown that forward current is higher than reverse current at the same voltage but the
difference gets smaller with increasing fluence. As a function of fluence the forward current is decreasing while the reverse current increasing.
Under reverse bias a rapid rise of current was observed above $\sim$ 600 V indicating significant charge multiplication effects.
Current increase with reverse bias was more pronounced at lower fluence indicating stronger multiplication factor. Annealing at 60$^\circ$C was studied. Small annealing effects of 
forward current was observed. Under reverse bias the current decreases with annealing time with a time dependence that can be approximately described with the annealing model and parameter values developed for results taken at low fluences.

Charge collection measurements performed with fast electrons from a $^{90}$Sr show that in this fluence range the collected charge increases linearly with
the reverse bias voltage up to 1000 V and above.  A mean collected charge of over 7000 electrons at 2.5$\times 10^{16}$ n/cm$^2$ and 3000 electrons at 10$^{17}$ n/cm$^2$ was measured.
A large increase of noise is observed with reverse bias values exceeding $\sim$ 700 V leading to a worse signal/noise ratio despite the larger collected charge.

In forward bias more charge than in reverse bias is collected at voltages below $\sim$ 400 V. High current is associated with larger noise so also at low bias values where the measured
signal is higher in forward bias, the signal/noise ratio is better in reverse bias.
It should be mentioned again at this point that signal/noise behaviour observed in this work could be significantly different than in tracking detector with much smaller pixel size so it only illustrates effects that govern the choice of optimal running regime.

Measurements of collected charge in forward bias were used to estimate effective trapping times at extreme fluences. Dependence of mobility on fluence measured in \cite{mikuz-trento}
was used and with this information trapping times were estimated from measurements of collected charge as a function of forward bias voltage. It was shown that trapping times are much shorter, at 10$^{17}$ n/cm$^2$ by a factor of more than 6, than values obtained by extrapolation from low fluence measurements. 

\section{Acknowledgment}
The authors would like to thank the crew of TRIGA mark II reactor for support with irradiation of samples and to acknowledge the financial support from the Slovenian Research Agency ( program ARRS P1-0135 and project ARRS J1-1699 ). This activity was partially supported by the Spanish Ministry of Science under grants RTI2018-094906-B-C22 and FPA2017-85155-C4-2-R; and the European Union Horizon 2020 Research and Innovation Framework Programme under Grant Agreement no. 654168 (AIDA-2020).


\begin{thebibliography}{99}

\bibitem{FCC_cdr}
A. Abada et al., {\it FCC-hh: the hadron collider}, The European Physical Journal Special Topics, vol. 228, no. 4, pp. 755 - 1107, Jul 2019., https://doi.org/10.1140/epjst/e2019-900087-0.

\bibitem{riegler}
W. Riegler, {\it  FCC-hh detector overview}, FCC week 2019, Brussels, Belgium, https://indico.cern.ch/event/727555/contributions/3461232/

\bibitem{radlevel}
ATLAS collaboration, I. Dawson, {\it Radiation background studies for the phase II inner tracker upgrade}, ATL-UPGRADE-PUB-2014-003, CERN, Geneva, Switzerland (2014).
s
\bibitem{pixelTDR}
ATLAS Collaboration, {\it Technical design report for the ATLAS inner tracker pixel detector}, CERN, Geneve, Switzerland, Tech. Rep. CERN-LHCC-2017-021; ATLAS-TDR-030, 2017.

\bibitem{spagheti}
G.. Kramberger et al., {\it Charge collection studies on custom silicon detectors irradiated up to 1.6E17 n$_{eq}$/cm$^2$},  2013 JINST 8 P08004.

\bibitem{mikuz-trento}
M. Miku\v z et al., {\it Electric field, mobility and trapping in Si detectors irradiated with neutrons and protons up to 10$^{17}$ n$_{eq}$/cm$^2$}, 11th ``Trento'' Workshop on Advanced Silicon Radiation
Detectors, Paris, 2016, https://indico.cern.ch/event/452766/contributions/1117366

\bibitem{mikuzFCC}
  M. Miku\v z et al., {\it Extreme Radiation Tolerant Sensor Technologies}, The 26th International Workshop on Vertex Detectors, 10-15 September 2017, Las Caldas, Spain,  https://indico.cern.ch/event/627245/contributions/2676707/

\bibitem{gianluigi_thin}
G. Casse et al., {\it Enhanced efficiency of segmented silicon detectors of different thicknesses after proton irradiations up to 1$\times 10^{16}$ n$_{eq}$ cm$^2$}, Nucl. Instrum. and Meth. A 624 (2010) 401.

\bibitem{cnm}
Institute of Microelectronics of Barcelona, IMB-CNM http://www.imb.cnm.csic.es/index.php/en

\bibitem{lgad}
G. Pellegrini et al., {\it Technology developments and first measurements of Low Gain Avalanche Detectors (LGAD) for high energy physics applications}, Nucl. Instrum. and Meth. A765 (2014).
 
\bibitem{carula}
  M. Carula et al., {\it Last measurements and developments on LGAD detectors}, 12th "Trento" Workshop on Advanced Silicon Radiation Detectors, Trento, Italy, 2017, https://indico.cern.ch/event/587631/contributions/2471730/
  
\bibitem{removal}
G. Kramberger et al., {\it Radiation effects in Low Gain Avalanche Detectors after hadron irradiations}, 2015 JINST 10 P07006.

 \bibitem{Reactor1}
L. Snoj, G. \v Zerovnik and A. Trkov, {\it Computational analysis of irradiation facilities at the JSI TRIGA
  reactor}, Appl. Radiat. Isot. 70 (2012) 483.

\bibitem{Reactor2}
K. Ambro\v zi\v c, G. \v Zerovnik and A. Trkov, {\it Computational analysis of dose rates at the JSI TRIGA
  reactor irradiation facilities}, Appl. Radiat. Isot. 130 (2017) 140.

\bibitem{Edge-TCT}
G. Kramberger et al., {\it Investigation of irradiated silicon detectors by Edge-TCT}, IEEE Trans. Nucl. Sci., Vol. 57 (2010) 2294.

\bibitem{Particulars}
{\it Particulars, Advanced measurement systems, Ltd.} www.particulars.si

\bibitem{paralel}
I. Mandi\' c et al., {\it Edge-TCT measurements with the laser beam directed parallel to the strips}, 2015 JINST 10 P08004.

\bibitem{scharf}
C. Scharf, F. Feindt, R. Klanner, {\it Influence of radiation damage on the absorption of near-infrared light in silicon},
arXiv:1905.03874v1, 2019.

\bibitem{SrSetup}  
G. Kramberger et al., {\it Charge collection properties of heavily irradiated epitaxial silicon detectors},
Nucl. Instrum. and Meth. A 554 (2005) 212.

\bibitem{field}
G. Kramberger, et al., {\it Modeling of electric field in silicon micro-strip detectors irradiated with neutrons and pions}, 2014 JINST 9 P10016.

\bibitem{klanner}
R. Klanner et al., {\it Determination of the electric field in highly-irradiated silicon sensors using edge-TCT measurements}, Nucl. Instrum. and Meth. A 951 (2020) 162987.

\bibitem{fielding}
  I. Mandi\` c et al., {\it Charge-collection efficiency of heavily irradiated silicon diodes operated with an increased free-carrier concentration and under forward bias}, Nucl. Instrum. and Meth. A 533 (2004) 442–453.

\bibitem{chilingarov}
A. Chilingarov, T. Sloan, {\it Operation of heavily irradiated silicon detectors under forward bias}, Nucl. Instrum. and Meth. A 399
(1997) 35.

\bibitem{ccestup}
G. Kramberger et al.,{\it Charge collection properties of heavily irradiated epitaxial silicon detectors},
Nucl. Instrum. and Meth. A 554 (2005) 212.
  
\bibitem{moll}
M. Moll, E. Fretwurst, G. Lindstr\"om, {\it Leakage current of hadron irradiated silicon detectors - material dependence}, Nucl. Instrum. and Meth. A 426 (1999) 87.

\bibitem{trap-krambi}
G. Kramberger et al., {\it Determination of effective trapping times for electrons and holes in irradiated silicon}, Nucl. Instrum. and Meth. A 476 (2002) 654.
   
\bibitem{trap-cindro}
V. Cindro et al., {\it Radiation damage in p-type silicon irradiated with neutrons and protons}, Nucl. Instrum. and Meth. 599(2009)65.

\bibitem{trap-CMS} 
T. Pholsen, Tracker group of the CMS collaboration,{\it  Trapping in proton irradiated p$^+$-n-n$^+$ silicon sensors at fluences anticipated at the HL-LHC outer tracker}, 2016 JINST 11 P04023.

\bibitem{irena_doc}
 I. Dolenc, {\it Development of Beam Conditions Monitor for the ATLAS experiment}, PhD thesis, University of Ljubljana, 2008, CERN-THESIS-2008-157.

\bibitem{jacoboni}
Jacoboni, C.Canali, G.Ottaviani and  A. Alberigi Quaranta, {\it A review of some charge transport properties of silicon}, Solid-State Electronics, Vol.20, 1977, p 77-89.

\bibitem{vsat}
R. Quay et al., {\it Temperature Dependent Model for the Saturation Velocity in Semiconductor Materials}, Materials Science in Semiconductor Processing, vol. 3, (2000) pp. 149-155.
  
\bibitem{jaztrento}
I. Mandi\' c et al., {\it Measurements with Si detectors irradiated to extreme fluences}, 15th ``Trento'' Workshop on Advanced Silicon Radiation
Detectors, Vienna, 2020, https://indico.cern.ch/event/813597/contributions/3727812/

\bibitem{DClighttrap}
G. Kramberger et al., {\it Determination of the effective dominant electron and hole trap in neutron-irradiated silicon detectors},
Nucl. Instrum. and Meth. A 516 (2004) 109.

\end{thebibliography}
\end{document}